\documentclass[final,5p,times,twocolumn]{elsarticle}
\biboptions{numbers,sort&compress}
\usepackage{graphicx}
\usepackage{float}
\usepackage{amsmath}
\usepackage[colorlinks,linkcolor=blue,citecolor=blue,urlcolor=black]{hyperref}
\usepackage{enumerate} 
\usepackage{amssymb}

\journal{Physics Letters B}

\begin{document}
\begin{frontmatter}
\title{Application of the microscopic optical potential of chiral effective field theory in astrophysical neutron-capture reactions}

\author[1]{Bing Wang}
\ead{bingw.phy@foxmail.com}
\address[1]{State Key Laboratory of Heavy Ion Science and Technology, Institute of Modern Physics, Chinese Academy of Sciences, Lanzhou 730000, China}
\author[2]{Dong Bai}
\ead{dbai@hhu.edu.cn}
\address[2]{College of Mechanics and Engineering Science, Hohai University, Nanjing 211100, China}
\author[3]{Yi Xu}
\ead{yi.xu@eli-np.ro}
\address[3]{Extreme Light Infrastructure - Nuclear Physics (ELI-NP), Horia Hulubei National Institute for R$\&$D in Physics and Nuclear Engineering (IFIN-HH), 077125 Buchurest-Magurele, Romania}

\begin{abstract}
A state-of-the-art microscopic global nucleon-nucleus optical potential has been developed by Whitehead, Lim, and Holt (WLH) within the framework of many-body perturbation theory, incorporating realistic nuclear interactions derived from chiral effective field theory. Given its potentially greater predictive power for reactions involving exotic isotopes, we apply it to the calculations of astrophysical neutron-capture reactions for the first time, which are particularly important to the nucleosynthesis of elements heavier than iron. It is found that this potential provides a good description of experimental known neutron-capture cross sections and Maxwellian-averaged cross sections. For unstable neutron-rich nuclei, we comprehensively calculate the neutron-capture reaction rates for all nuclei with $26\leq Z\leq84$, located between the valley of stability and the neutron drip line, using the backward-forward Monte Carlo method with the $f_{rms}$ deviation as the $\chi^2$ estimator. The results reveal a noticeable separation in the uncertainty of rates around an isospin asymmetry of 0.28 under the constraint $f_{rms} \leq 1.56$. This highlights the critical role of isospin dependence in optical potentials and suggests that future developments of the WLH potential may pay special attention to the isospin dependence.
\end{abstract}


\end{frontmatter}

\section{Introduction}
The elements heavier than iron in the Universe are mostly synthesized through neutron-capture reactions \cite{burbidge1957,wallerstein1997a}. The slow ($s$) \cite{busso1999,pignatari2010,kappeler2011,lugaro2023} and rapid ($r$) \cite{arnould2007,thielemann2011,kajino2019,cowan2021} neutron-capture processes with their neutron-capture timescales slower or more rapid, respectively, than $\beta$ decay timescale of unstable nuclei are responsible for the vast majority of heavy element abundances. Between these two processes, other neutron-capture processes, such as the intermediate ($i$) \cite{hampel2016,denissenkov2017,choplin2021} and neutron ($n$) \cite{meyer2000,pignatari2018} processes, may also be indispensable for explaining certain specific features in elemental abundance patterns. However, a large number of neutron-capture rates involved in these nucleosynthesis processes are not known experimentally \cite{goriely1998,goriely2021,choplin2023}, and only theoretical predictions can fill the gaps. Theoretical neutron-capture rates are commonly treated within the framework of the Hauser-Feshbach statistical model \cite{hauser1952,goriely2008}. The optical potential, nuclear level density, and $\gamma$-ray strength function are crucial nuclear structure inputs to this model. Significant efforts have been devoted to improving the descriptions of the nuclear level densities, both from phenomenological \cite{gilbert1965,dilg1973,Ignatyuk1979,wang2019,wang2020} and microscopic \cite{demetriou2001,hilaire2006,goriely2008a} perspectives, and the strength functions through phenomenological \cite{Brink1957,axel1962,kopecky1990,goriely2019c,wang2023a} and microscopic \cite{goriely2002,goriely2004,xu2021} approaches. 

For the optical potential, which is essential for calculating the neutron transmission coefficient, a large number of developed models are available \cite{capote2009a}. These include phenomenological potentials \cite{koning2003}, semi-microscopic potentials \cite{bauge1998a,bauge2000a,bauge2001a}, dispersive potentials \cite{morillon2007a}, as well as models with uncertainty quantification \cite{beyer2025,pruitt2024a,pruitt2023b}. However, most of them either rely on fitting nuclear reaction data, introducing unreliability when extrapolated to exotic nuclei lacking experimental constraints, or have limited applicability ranges, preventing their global use in astrophysical neutron-capture reactions \cite{hebborn2023a}. The microscopic optical potentials developed from chiral effective field theory is a new booming area \cite{durant2022c,durant2022b,bai2021,durant2020,durant2018}. Recently, Whitehead, Lim, and Holt (WLH) constructed the full microscopic global nucleon-nucleus optical potential based on an analysis of 1800 isotopes within the framework of many-body perturbation theory using realistic nuclear interactions from chiral effective field theory \cite{whitehead2021}. This state-of-the-art potential, derived from the chiral nuclear force without any parameters fitted to nucleon-nucleus scattering data, is expected to be more powerful in predicting neutron-capture rates for nuclei far from stability. Moreover, it provides quantified uncertainties for the parameters, enabling one to estimate the uncertainty in a given reaction observable. While the WLH potential displays impressive capability in describing elastic scattering \cite{whitehead2021}, charge-exchange reactions \cite{whitehead2022}, and transfer reactions \cite{xu2024}, it has not yet been widely applied to astrophysical neutron-capture reactions.

In this work, we apply the WLH potential to the study of astrophysical neutron-capture reactions for the first time and test its applicability to nuclei with $26\leq Z\leq84$ lying between the neutron drip line and the valley of $\beta$ stability, which is a critical region for the nucleosynthesis of elements heavier than iron. The rest of this article is organized as follows. In Sec. \ref{sec:thmo}, the WLH potential model used in our work is described. In Sec. \ref{sec:resdis}, the calculated results are compared with experimental data, and the WLH potential is applied to neutron-rich nuclei. In Sec. \ref{sec:con}, the conclusions are presented.

\section{WLH potential model}
\label{sec:thmo}
In the WLH potential, a global optical potential is parameterized for each of the five chiral interactions with a Woods-Saxon form \cite{whitehead2021},
\begin{align}
	U(r,E)=&-\mathcal{U}_Vf(r;r_V,a_V)-i\mathcal{U}_Wf(r;r_W,a_W) \notag \\
	       &+i4a_S\mathcal{U}_S\frac{d}{dr}f(r;r_S,a_S)  \notag \\
	       &+\mathcal{U}_{SO}\frac{1}{m^2_{\pi}}\frac{1}{r}\frac{d}{dr}f(r;r_{SO},a_{SO})\vec{\ell}\cdot\vec{\sigma}.
\end{align}
It consists of a real volume term $U_V$, an imaginary volume term $U_W$, an imaginary surface term $U_S$, and a real spin–orbit term $U_{SO}$. The Woods-Saxon shape factor is given by 
\begin{align}
	f(r;r_i,a_i)=\frac{1}{1+e^{(r-A^{1/3}r_i)/a_i}},
\end{align}
where $A$ is the mass number, and $r_i$ and $a_i$ are the geometry parameters .The well depths ($\mathcal{U}_V$, $\mathcal{U}_W$, $\mathcal{U}_S$, and $\mathcal{U}_{SO}$) and the geometry parameters depend smoothly on $A$, the projectile energy, and the isospin asymmetry, with only a few parameters. The functional forms of these parameters can be found in Ref. \cite{whitehead2021}. \\
\indent The dominant source of theoretical uncertainty in the WLH potential arises from the choice of chiral potential. Therefore, in Ref. \cite{whitehead2021}, the covariance matrix for all the global optical potential parameters derived from five chiral interactions, along with their mean values, are used to generate a multivariate normal distribution that includes the correlations between parameters. Random parameter sets for the global optical potential can then be produced by sampling from this distribution. This enables one to estimate the uncertainty in a given reaction observable through many samples of the global optical potential. As a result, the multivariate distribution, and the uncertainty obtained from many samples within it, are independent of any experimental nucleon-nucleus reaction data, and the theoretical uncertainty stems directly from the uncertainty in the fundamental chiral interaction. \\
\indent In our calculations, we sample parameter sets from the multivariate distribution which is implemented in Python code of Ref. \cite{whitehead2021}. The uncertainty range used for sampling is consistent with the original uncertainty ranges of the WLH parameters. The corresponding reaction observables for each sample are computed within the framework of the Hauser-Feshbach statistical model using the reaction code TALYS \cite{koning2023}.

\section{Calculations and results}
\label{sec:resdis}
\subsection{Comparison with experimental result}
\label{sec:cexp}
\begin{figure*}
	\centering
	\includegraphics[width=1\linewidth]{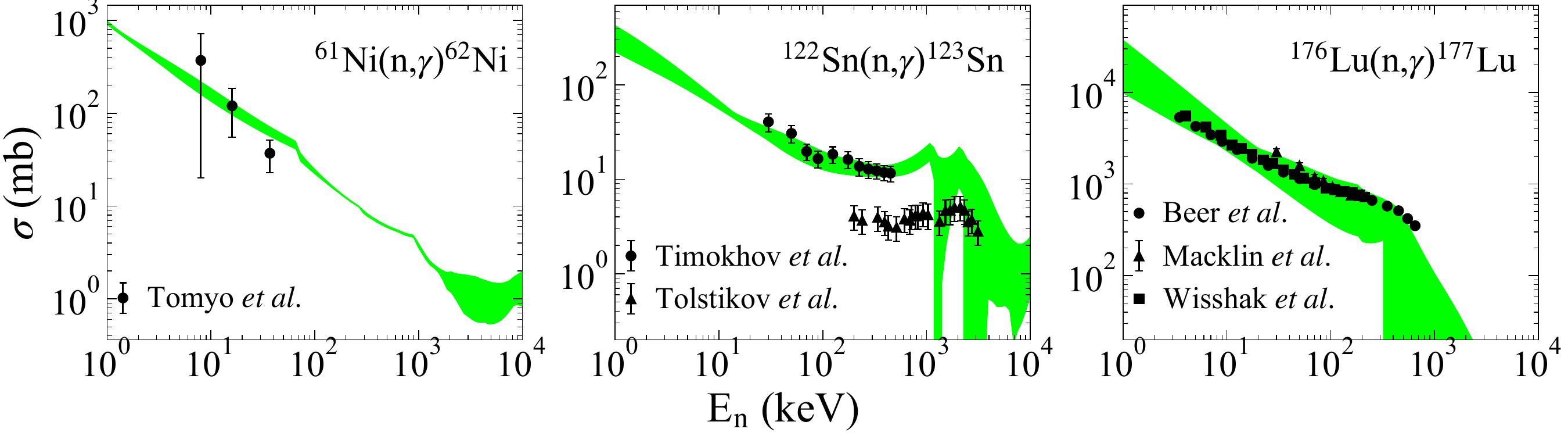}
	\caption{Comparison of the calculated and experimental neutron-capture cross sections as a function of incident energy $E_n$ for three nuclei. The shaded area represents cross sections calculated from 400 random samples of the WLH potential. Experimental data are shown as black points and are taken from Refs. \cite{tomyo2005} for $^{61}\text{Ni}(\text{n},\gamma)^{62}\text{Ni}$, \cite{tolstikov1968,timokhov1988} for $^{122}\text{Sn}(\text{n},\gamma)^{123}\text{Sn}$, and \cite{beer1984,macklin1967,wisshak2006} for $^{176}\text{Lu}(\text{n},\gamma)^{177}\text{Lu}$.}
	\label{cross_sec}
\end{figure*}

In order to test the capability of the WLH potential in describing the cross sections of neutron-capture reactions, we compare in Fig. \ref{cross_sec} our calculated results, obtained by sampling the potential parameter set 400 times, with experimental data for the nuclei $^{61}\text{Ni}$, $^{122}\text{Sn}$, and $^{176}\text{Lu}$. A rather good agreement is found, particularly in the low-energy region, with the experimental data primarily lying within the calculated uncertainty ranges, which is especially significant for astrophysical nucleosynthesis.

\begin{figure}
	\centering
	\includegraphics[width=1\linewidth]{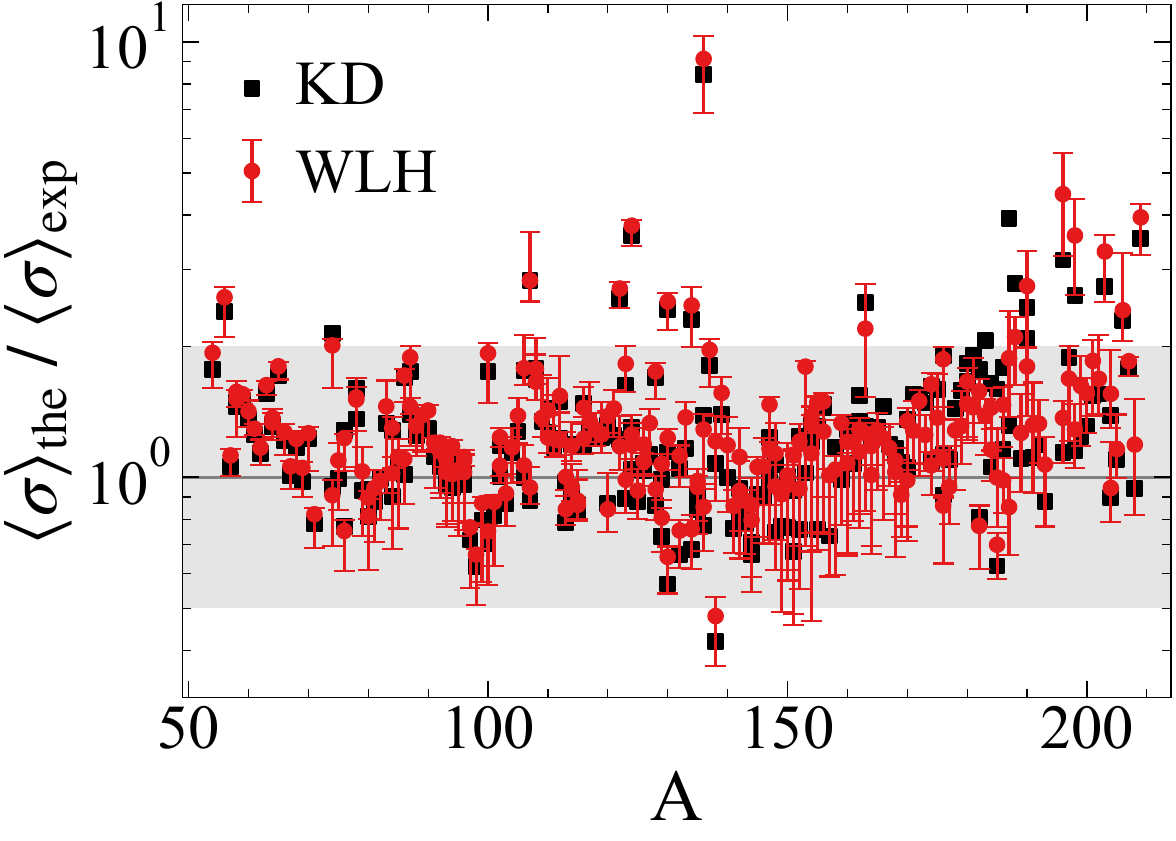}
	\caption{Ratios of theoretical to experimental ($n,\gamma$) MACS for the 221 ($26\leq Z\leq 83$) known KADoNiS values at 30 keV. Black squares are calculated using the phenomenological KD potential. Red circles represent the results with the WLH potential, with error bars indicating the maximum and minimum values under the constraint $f_{rms}\leq 1.56$. The gray shaded area guides the eye for deviations within a factor of two.}
	\label{MACS_ratio}
\end{figure}

To further assess the global predictive ability of the WLH potential against all available experimental data, we calculate the Maxwellian-averaged cross sections (MACS) at 30 keV for 221 nuclei with $26\leq Z\leq 83$, and compare the results with experimental data taken from the KADoNiS database \cite{dillmann2006a}. The average values of the WLH potential parameter sets are used in the calculations. Other model parameters, such as those for level density and photon strength functions, are taken, wherever possible, from the recommended values in Ref. \cite{rochman2025}, which are suggested for TALYS calculations in astrophysics applications. As shown by the red circles in Fig. \ref{MACS_ratio}, the deviations between the MACS values calculated using the WLH potentials and those reported in KADoNiS are generally within a factor of two. Furthermore, we present the theoretical results obtained using the phenomenological potential of Koning and Delaroche (KD) which is widely employed in nuclear reaction calculations and is well known for its good agreement with experimental data through parameter adjustment \cite{koning2003}. It can be seen in Fig. \ref{MACS_ratio} that the results based on the WLH potentials are consistent with those obtained from the KD potentials in both magnitude and overall trend. The consistency of the WLH results with the experimental MACS data, as well as with the KD results, demonstrates that the WLH potentials are capable of reproducing experimental neutron-capture data, though not with greater accuracy than KD.

\begin{figure}
	\centering
	\includegraphics[width=1\linewidth]{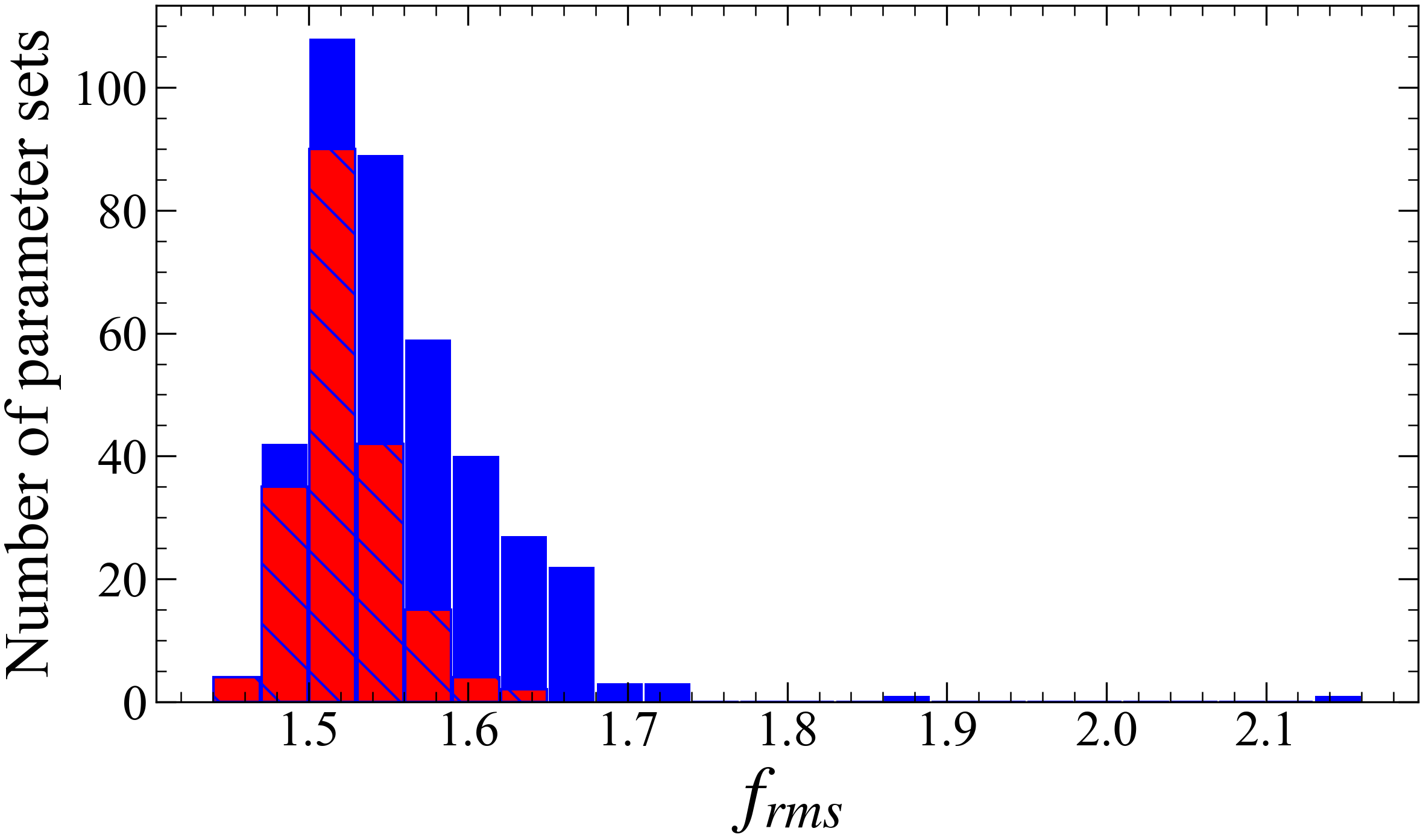}
	\caption{Distribution of $f_{rms}$ deviations. Blue bars show the deviations with respect to the 221 experimental ($n,\gamma$) MACS for 400 sampled parameter sets. Red bars correspond to the cases with $f'_{frms}\leq 2.1$ with respect to the 298 experimental neutron strength functions.}
	\label{frms_hist}
\end{figure}

Sampling of the potential parameter set introduces an uncertainty range in the calculated results. To quantify the predictive accuracy of the WLH potential for each sample, we adopt the root-mean-square (rms) criterion, i.e. $f_{rms}$ deviation, defined in Ref. \cite{rochman2025,martinet2024} as
\begin{align}
	f_{rms}=\exp\left[\frac{1}{N_e}\sum_{i=1}^{N_e}\ln^2\left(\frac{\left\langle\sigma_{\mathrm{th},i}\right\rangle}{\left\langle\sigma_{\exp,i}\right\rangle}\right)\right]^{1/2},
	\label{frms}
\end{align}
where $N_e$ is the number of known MACS from KADoNiS database, and $\left\langle \sigma_{\mathrm{th},i} \right\rangle$ and $\left\langle \sigma_{\exp,i} \right\rangle$ are the theoretical and experimentally MACS, respectively. With Eq. \eqref{frms}, each sampling corresponds to an $f_{rms}$ deviation value, which is always greater than or equal to 1. The deviation from 1 indicates, on average, how far the prediction deviates from the experimental data. In Fig. \ref{frms_hist}, the distribution of $f_{rms}$ deviations with respect to the 221 experimental ($n,\gamma$) MACS at 30 keV from the KADoNiS database is displayed. As can be seen, while the $f_{rms}$ deviations reach up to 2.16, most are below 1.74. 
Even though the $f_{rms}$ deviations are generally small, they are consistent with those reported in Refs. \cite{rochman2025,martinet2024}, which are based on uncertainty analyses of neutron-capture reactions relevant to astrophysics. Smaller $f_{rms}$ values mean that the calculated MACS is closer to the experimental data and further confirm the capability of the WLH potentials to reproduce neutron-capture experimental data. In other words, the small $f_{rms}$ values also support the fact that the low-energy neutron-capture cross sections of stable nuclei are not highly sensitive to the optical potentials.

\begin{figure*}
	\centering
	\includegraphics[width=1\linewidth]{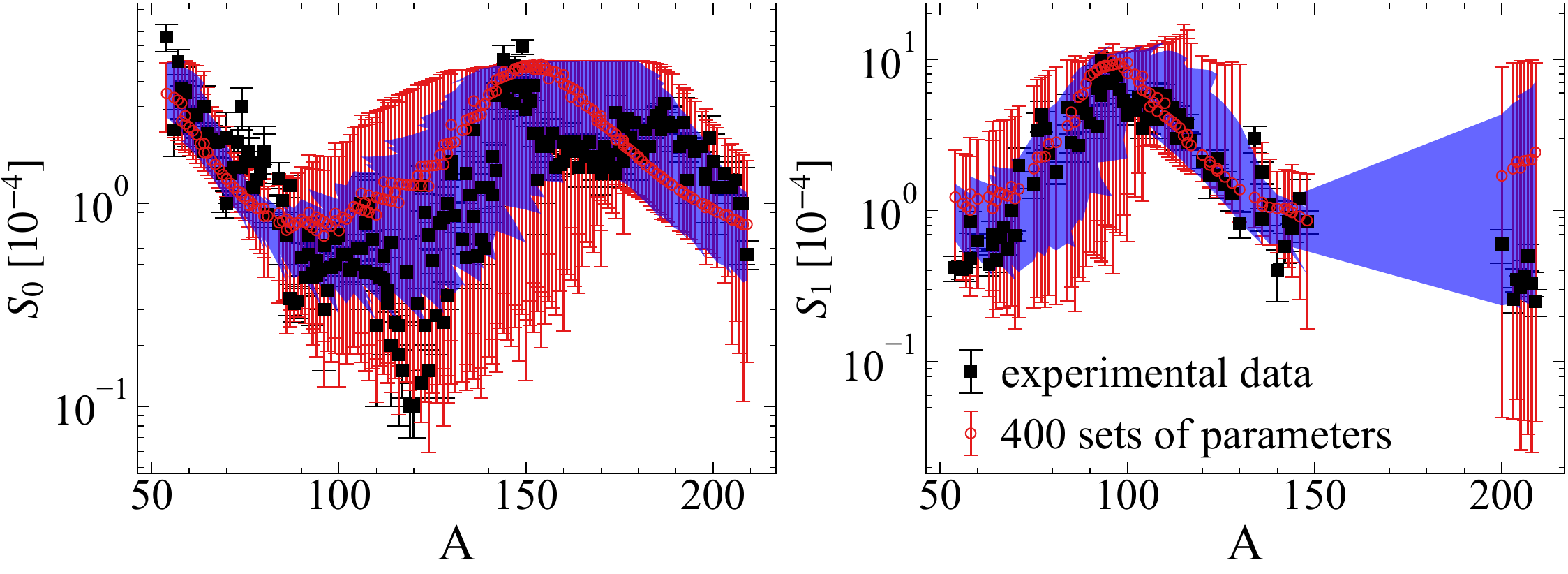}
	\caption{Comparison between the experimental $S$- and $P$-wave strength functions ($S_0$ and $S_1$) for 298 nuclei ($26\leq Z\leq 83$) and the theoretical predictions. The red circles show the values calculated using 400 sets of WLH potential parameters. The blue shaded areas correspond to the parameter sets constrained by $f_{rms}\leq 1.56$.}
	\label{s0s1}
\end{figure*}

The sensitivity of the neutron strength function to optical potentials is investigated as well. Using the 400 sets of parameters, we calculate the neutron strength functions for the S-wave ($S_0$) of 207 nuclei and the P-wave ($S_1$) of 91 nuclei, and compare them with the experimental data taken from RIPL-3 \cite{capote2009b}, as shown in Fig. \ref{s0s1}. It can be seen that the WLH potentials are globally compatible with the neutron strength data, but they span a much larger variation in the predicted strengths, which could be considered for further constraining the WLH potentials.

\subsection{Application to neutron-rich nuclei}
\label{sec:nrich}
Applying the WLH potential to neutron-capture reactions of nuclei far from stability, for which no experimental data are available, leads to increasingly intensive computations. This is due to the large number of neutron-rich nuclei involved, each requiring extensive sampling of potential parameters across a wide uncertainty range, along with the corresponding reaction calculations. For the purpose of reducing computational cost while efficiently propagating parameter uncertainties to the observables, as well as to investigate the properties of the WLH potential under different parameter constraint conditions, we adopt the backward-forward Monte Carlo (BFMC) method \cite{martinet2024,chadwick2007,bauge2011,goriely2014}. This method relies on the sampling of the model parameters and the use of a generalized $\chi^2$ estimator to quantify the likelihood of each simulation with respect to a given set of experimental constraints. It consists primarily of two Monte Carlo (MC) steps: in the backward MC step, suitable parameter samples are selected based on experimental constraints; in the forward MC step, the selected parameter samples are used to calculate unmeasured quantities.

We use the $f_{rms}$ deviation as the $\chi^2$ estimator with respect to the 221 experimental MACS values in the backward MC step. Specifically, we apply a threshold function
\begin{align}
	w=\left\{
	\begin{aligned}
		&1\quad\mathrm{for}\quad f_{rms}\leq f_{rms}^{crit} ,\\
		&0\quad\mathrm{for}\quad f_{rms}> f_{rms}^{crit} ,
	\end{aligned}
	\right.
\end{align}
to select the sampled potential parameter sets for which the $f_{rms}$ deviation is less than or equal to the critical value $f_{rms}^{crit}$. This ensures that the corresponding calculated observables are in closer agreement with the experimental data than those with $f_{rms}> f_{rms}^{crit}$. In Fig. \ref{MACS_ratio}, the error bars of the WLH potentials are derived from potential parameters constrained by $f_{rms}\leq 1.56$. It can be seen that the uncertainty range of the WLH potentials effectively encompasses the theoretical results of the KD potentials.

\begin{figure*}
	\centering
	\includegraphics[width=1\linewidth]{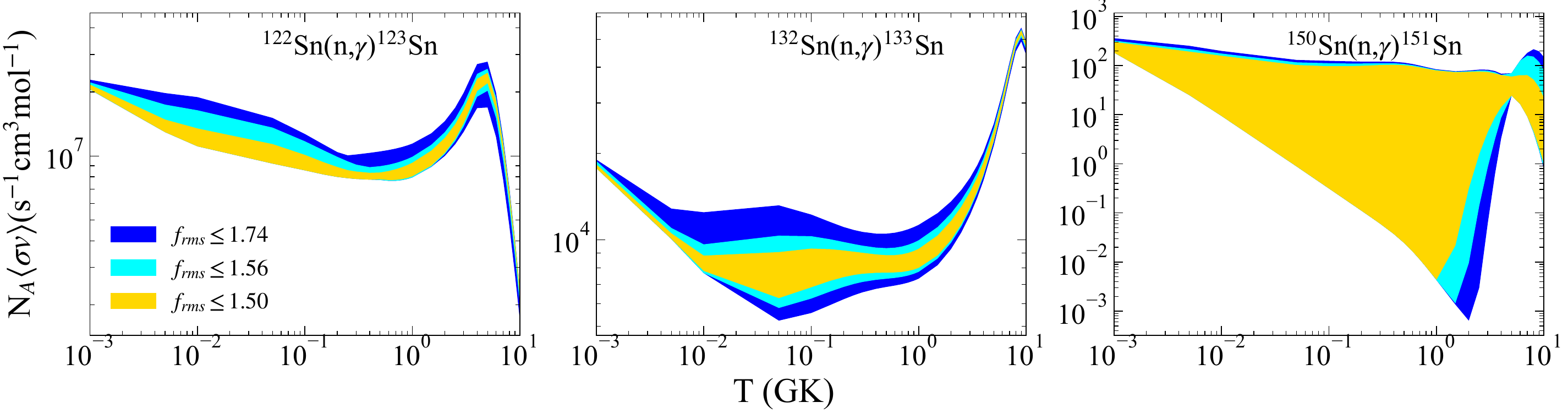}
	\caption{The neutron-capture rates, constrained by three $f_{rms}^{crit}$ values (1.50, 1.56 and 1.74) corresponding to three differently shaded areas, as a function of temperature for the isotopes $^{122}\text{Sn}$, $^{132}\text{Sn}$, and $^{150}\text{Sn}$.}
	\label{Sn_Ts}
\end{figure*}

\begin{figure}
	\centering
	\includegraphics[width=1\linewidth]{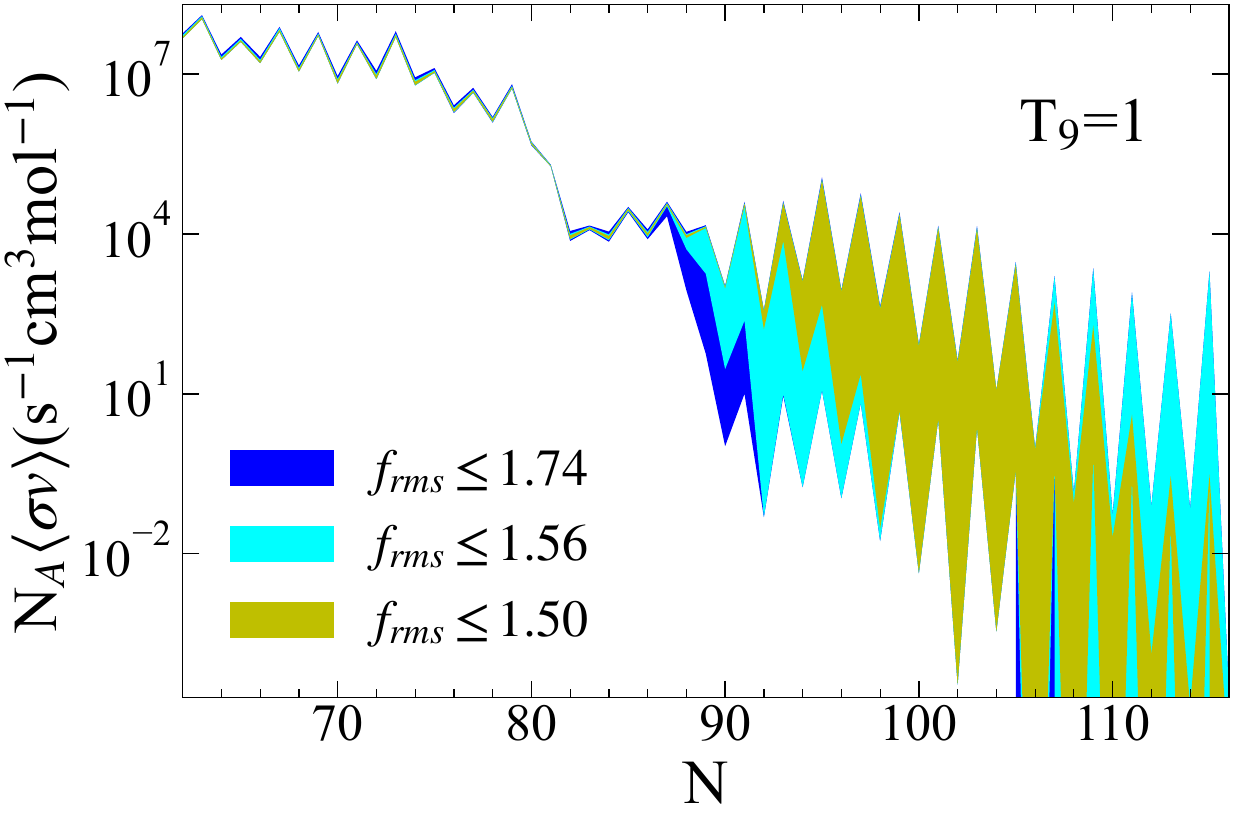}
	\caption{The neutron-capture rates for Sn isotopes as a function of neutron number $N$ at a temperature of $T=1$ GK. Differently shaded areas represent different $f_{rms}^{crit}$ values.}
	\label{Sn_T1}
\end{figure}

In the forward MC step, the parameter sets selected in the backward MC step are used to calculate the neutron-capture rates of neutron-rich nuclei. As shown in Fig. \ref{Sn_Ts}, the neutron-capture rates and their uncertainty ranges as a function of temperature for the Sn isotopes $^{122}\text{Sn}$, $^{132}\text{Sn}$, and $^{150}\text{Sn}$ are calculated using three $f_{rms}^{crit}$ values. As $f_{rms}^{crit}$ decreases from 1.74 to 1.50, the uncertainty ranges clearly decrease, indicating that it is valid to use the experimental MACS to constrain the selection of parameter sets in the backward MC step, and that parameter sets closer to the experimental data are preferentially selected. For $^{122}\text{Sn}$ and $^{132}\text{Sn}$, the uncertainty ranges are within one order of magnitude. For $^{150}\text{Sn}$, however, the differences in the reaction rates reach up to five orders of magnitude. Fig. \ref{Sn_T1} shows the neutron-capture rates of Sn isotopes as a function of neutron number $N$ at a temperature of $T=1$ GK. It is more clearly seen that the uncertainty in the rates decreases with decreasing $f_{rms}^{crit}$, and when $N$ exceeds approximately 87, the uncertainty increases dramatically and exhibits an odd-even effect. 

\begin{figure*}
	\centering
	\includegraphics[width=1\linewidth]{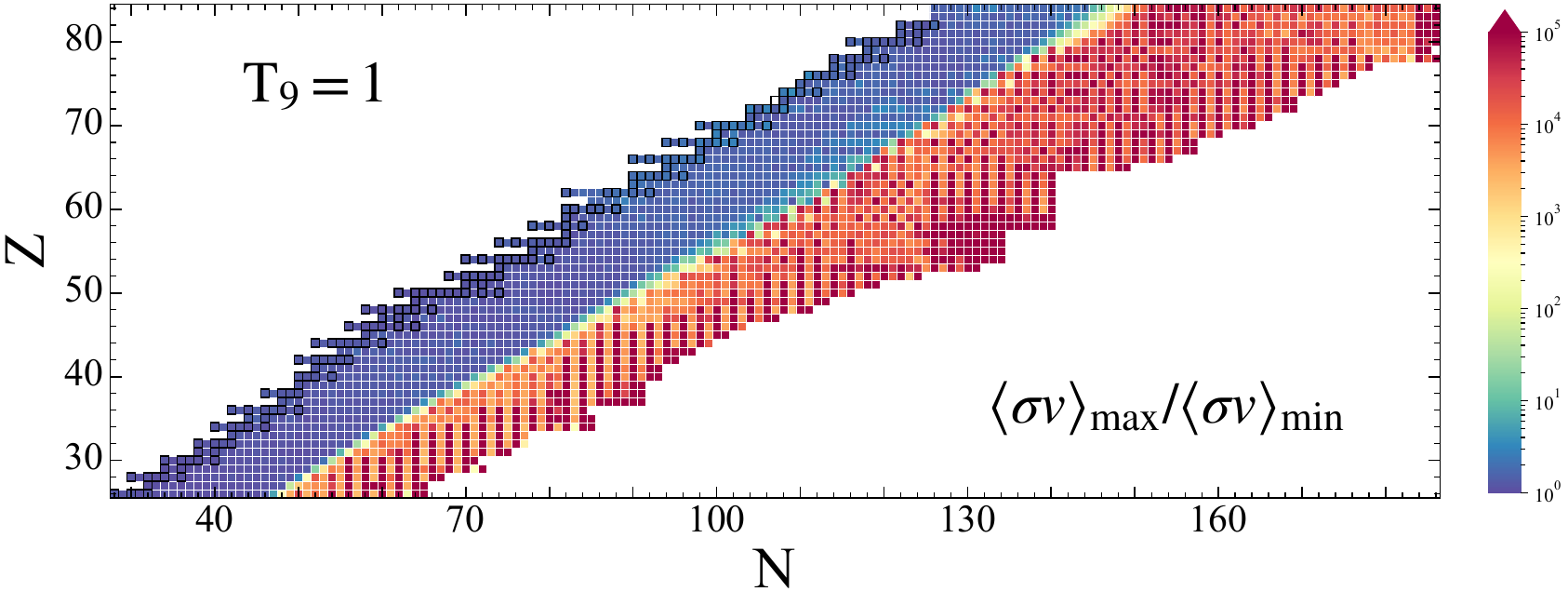}
	\caption{Representations in the ($N,Z$) plane of the ratio between the maximum and minimum reaction rates for all nuclei with $26\leq Z\leq84$, lying between the valley of stability and neutron drip line, at $T=1$ GK and $f_{rms}\leq 1.56$.}
	\label{chart_t1}
\end{figure*}

Furthermore, using the parameter sets constrained by $f_{rms}\leq 1.56$, we comprehensively calculate the neutron-capture rates at $T=1$ GK for all nuclei with $26\leq Z\leq84$, located between the valley of stability and the neutron drip line. The ratio of maximum to minimum rates is shown in the chart of the nuclides in Fig. \ref{chart_t1}. There is a noticeable separation around an isospin asymmetry, i.e., $\delta=(N-Z)/A$, of 0.28, below which the uncertainty range in the reaction rate remains within one order of magnitude, and above which it mostly exceeds three orders of magnitude. 
In addition, Fig. \ref{Sn_T1} for the Sn isotopes shows that the separation of uncertainties related to isospin asymmetry always occurs under the constraints of $f_{rms}^{crit}$ from 1.50 to 1.74, especially at 1.74, where almost no constraint is imposed. This implies that these separations are a general feature of the WLH potential and do not result from the constraint conditions. A significant effect of the isospin asymmetry on the astrophysical neutron-capture reactions, especially for nuclei off the stability line, has already been suggested \cite{goriely2007a} and recently corroborated by experimental evidence \cite{voinov2021}. As pointed out in Ref. \cite{goriely2007a}, the isovector imaginary component of the optical potential has a drastic impact on the neutron capture cross section for neutron-rich nuclei. Similarly, we believe that the separations observed in Figs. \ref{Sn_T1} and \ref{chart_t1} may also be driven by the imaginary component associated with the isospin asymmetry in the WLH potential. For instance, from the expression of the imaginary volume term $U_W$ given in Ref. \cite{whitehead2021}, it can be seen that with increasing isospin asymmetry, $U_W$ may even take on negative values. This could be one of the factors responsible for the sharp increase in the uncertainties. 

Moreover, the calculation results for the strength functions obtained with the parameter sets constrained by $f_{rms}\leq 1.56$ are shown in Fig. \ref{s0s1}, which exhibit improved predictions with reduced uncertainties. We also tentatively calculate the $f'_{rms}$ deviations with respect to the 298 experimental neutron strength functions and find that nearly half of the parameter sets have $f'_{rms}$ values below 2.1. The $f_{rms}$ distributions corresponding to the parameter sets with $f'_{rms}\leq 2.1$ are displayed in Fig. \ref{frms_hist}. It can be seen that the $f_{rms}$ values associated with $f'_{rms}\leq 2.1$ predominantly lie within the range $f_{rms}\leq 1.56$. These consistent constraints from the MACS and the strength functions indicate that similar conclusions to those presented in Figs. \ref{Sn_Ts}--\ref{chart_t1} would be reached when using either constraint. Ideally, employing the BFMC method with simultaneous constraints from both observables, MACS and strength functions, would be worthwhile in future work.

\section{Conclusions}
\label{sec:con}
In this work, we apply the microscopic global optical potential of chiral effective field theory (the WLH potential) to the study of astrophysical neutron-capture reaction for the first time. By comparing the neutron-capture cross sections calculated using the WLH potentials with experimental data for the nuclei $^{61}\text{Ni}$, $^{122}\text{Sn}$, and $^{176}\text{Lu}$, we find that the theoretical predictions show good agreement with the experimental results. Furthermore, for all experimentally known MACS with $26\leq Z\leq 83$, the WLH potentials also provide a satisfactory description. We employ the $f_{rms}$ deviation to assess the predictive accuracy of the WLH potential for each sample of the parameter set, and find that the associated uncertainty lies within a reasonable range. These results demonstrate the capability of the WLH potentials to reliably reproduce experimental data and establish a foundation for its application to unstable neutron-rich nuclei.

For neutron-rich nuclei, to efficiently propagate parameter uncertainties to reaction rates and reduce computational cost, we adopt the BFMC method, using the $f_{rms}$ deviation to constrain parameter sampling. With this approach, we calculate the neutron-capture rates for nuclei $^{122}\text{Sn}$, $^{132}\text{Sn}$, and $^{150}\text{Sn}$ at different temperatures, as well as the rates for Sn isotopes as a function of neutron number at a temperature of $T=1$ GK. It is found that the $f_{rms}$ constraint effectively controls the propagation of uncertainty, and a sharp increase in uncertainty is observed around $N=87$. A further comprehensive calculation, covering all nuclei with $26\leq Z\leq84$ between the valley of stability and the neutron drip line at $T=1$ GK and constrained by $f_{rms}\leq 1.56$, shows that the WLH potentials provide a good description with relatively small uncertainties for nuclei with isospin asymmetry $\delta <0.28$. However, for nuclei with $\delta>0.28$, the uncertainty in the reaction rates increases significantly. This highlights the critical role of isospin dependence in optical potentials and suggests that future developments of the WLH potential may pay greater attention to this aspect.

The MACS of neutron capture is directly related to the astrophysical cross section, and a large amount of experimental data can be utilized. 
Other experimental observables, such as elastic scattering data and the neutron strength function, may be more sensitive to the neutron optical potentials and thus offer stronger constraints to the parameter sampling. Following the sensitivity check of neutron optical potentials to the neutron strength function presented in this work, a systematic quantitative study about how the neutron optical potentials can be constrained by different experimental observables would be a worthwhile direction in future.

Meanwhile, in the WLH potential, nuclear deformation is neglected, and its effects are not included in the present study. However, since nuclear deformation plays an important role in reproducing nucleon–nucleus scattering experimental data, incorporating this effect into the WLH potential to describe neutron-capture reactions would be worthwhile in future work.

\section{Acknowledgements}
This work is supported by the National Key Research and Development program (MOST 2022YFA1602304), the National Natural Science Foundation of China (Grants No.\ 12335009, 12175156, and 12375122),  the Strategic Priority Research Program of Chinese Academy of Sciences (Grant No.\ XDB34020200), and the Fundamental Research Funds for the Central Universities (Grant No.\ B240201048). Y. X. acknowledges support from the Romanian Ministry of Research, Innovation and Digitization, CNCS-UEFIS-CDI, project number PN-IV-P1-PCE-2023-0384 within PNCDI IV, the Romanian Ministry of Research, Innovation and Digitalization under Contract No. PN 23 21 01 06, the ELI-RO project with Contract ELI-RORDI-2024-008 (AMAP), the IAEA Coordinated Research Project on Updating Nuclear Level Densities for Applications (F41034) under Contract No. 28638.

\bibliographystyle{elsarticle-num}
\bibliography{cited.bib}

\end{document}